# Limits on direct detection of neutralino dark matter from b → sγ decays

Lars Bergström

*Stockholm University, Department of Physics,*
*Box 6730, S-113 85 Stockholm, Sweden*
*and*
*University of Uppsala, Department of Theoretical Physics,*
*Box 803, S-751 08 Uppsala, Sweden*

and

Paolo Gondolo

*Université de Paris VII, Physique Théorique et Hautes Energies,*
*Tour 14-24, 5$^e$ étage, 2 place Jussieu, 75005 Paris, France*
*and*
*University of Oxford, Department of Physics, Theoretical Physics,*
*1 Keble Road, Oxford, OX1 3NP, United Kingdom*

## Abstract

We analyze the rate of detection of minimal supersymmetric neutralino dark matter in germanium, sapphire and sodium iodide detectors, imposing cosmological and recent accelerator bounds including those from b → sγ decay. We find, in contrast with several other recent analyses, that although the b → sγ constraint reduces the number of viable models, models still remain where the counting rate in solid state detectors exceeds 10 kg$^{-1}$ day$^{-1}$.

The recent observation by the CLEO collaboration of the b → sγ decay [1] has stirred interest in the possible bounds obtainable for supersymmetric models that contribute to this process [2]. Some authors [3, 4] have gone one step further and analyzed the consequences of this and other accelerator bounds for the predicted rates in experiments aimed at detecting neutralino dark matter. One problem with even the so-called "minimal" supersymmetric standard model (MSSM) is that it contains a large number of parameters which in principle are free, although several of them are constrained by various naturalness arguments and experimental bounds on flavor changing interactions etc. In order to reduce the number of parameters, some assumptions related to supersymmetric grand unification of coupling constants and masses are usually made. Even more restricted models have been suggested (e.g., [5]), where the simplest possible structure is chosen at the GUT scale and the electroweak symmetry breaking is assumed to be achieved only through radiative corrections to the Higgs potential when the GUT parameters are run to lower energies using the renormalization group equations.

As expected, making the model more restrictive gives more restrictive bounds on the experimental signals of the lightest supersymmetric particles, both in accelerator experiments and in dark matter detection experiments (if supersymmetry is related at all to the dark matter problem of the galactic halo). In particular, it was shown in [3, 4, 5] that rates for both direct and indirect detection of neutralino dark matter generally become very small in this type of models. At this point it should be realized, however, that evidence is mounting that whatever the effective low-energy supersymmetric model is, it is unlikely to be of such a simplified form. For instance, in models derived from superstrings it is not clear if there is an intermediate GUT scale at all. Even if there is, threshold corrections at the GUT scale may be important (for a recent discussion of some of these problems, see [6]). In fact, recent analyses of the MSSM phenomenology have appeared in which the universality condition of the scalar masses at the GUT scale is relaxed, especially for the Higgs sector [7]. It has also recently been suggested [8] that to reproduce the measured value of $\alpha_s$ at low energies, it may even be necessary to modify one of most frequently used relations, that of the gaugino mass parameters (we follow the notational convention of [9]):

$$
\begin{aligned}
M_1 &= \tfrac{5}{3} \tan^2 \theta_w M_2 \simeq 0.5 M_2, \\
M_2 &= \frac{\alpha_{ew}}{\sin^2 \theta_w \alpha_s} M_3 \simeq 0.3 M_3.
\end{aligned}
\qquad (1)
$$

The upshot of this is that although the simplistic models have the virtue of being predictive, and due to small rates largely irrelevant for present-day dark matter detectors, the price paid is the lack of generality.

If one adopts a phenomenological approach and allows for a more general variation of parameters in the MSSM, still of course consistent with experimental bounds and giving correct low-energy symmetry breaking, one may be tempted to define a probability measure in parameter space. That is, one may want to translate the number of allowed models giving, e.g., a range of detection rates to a "probability". This method is often used implicitly as one typically scans a large number of models and presents results as density of points in some parameter plane. It may then be tempting to assign values where there is a large density of model points a higher probability than those with small density. We will argue, illustrated by some simple examples, that this is not meaningful to do, since different parametrizations of the models can give completely different results.

The conservative approach we propose is to regard the whole range of outcomes of a calculation as a priori equally probable, irrespective of the parametrization. This means that really only upper and lower limits can be given. Another consequence of this philosophy is that it becomes important to scan the model parameter space accurately enough to discover also the very extreme outcomes of the calculation, since they will be used for setting the bounds.

We started this study with the intention of relaxing the various GUT or radiative breaking conditions to explore the full range of possibilities in a generic minimal supersymmetric model. We found, to our surprise, that even without relaxing those assumptions there still were models in which the dark matter detection rates in solid state detectors were quite high. We think that this discrepancy with published results may be due to insufficient scanning of parameters space in the earlier works.

## The minimal supersymmetric standard models

We work in the framework of the minimal $N = 1$ supersymmetric extension of the standard model defined by, besides the particle content and gauge couplings required by supersymmetry, the superpotential

$$W = \epsilon_{ij} \left( -\hat{\mathbf{e}}_R^* \mathbf{Y}_E \hat{\mathbf{l}}_L^i \hat{H}_1^j - \hat{\mathbf{d}}_R^* \mathbf{Y}_D \hat{\mathbf{q}}_L^i \hat{H}_1^j + \hat{\mathbf{u}}_R^* \mathbf{Y}_U \hat{\mathbf{q}}_L^i \hat{H}_2^j - \mu \hat{H}_1^i \hat{H}_2^j \right) \tag{2}$$

and the soft supersymmetry-breaking potential

$$\begin{aligned} V_{\text{soft}} &= \epsilon_{ij} \left( -\tilde{\mathbf{e}}_R^* \mathbf{A}_E \mathbf{Y}_E \tilde{\mathbf{l}}_L^i H_1^j - \tilde{\mathbf{d}}_R^* \mathbf{A}_D \mathbf{Y}_D \tilde{\mathbf{q}}_L^i H_1^j + \tilde{\mathbf{u}}_R^* \mathbf{A}_U \mathbf{Y}_U \tilde{\mathbf{q}}_L^i H_2^j - B\mu H_1^i H_2^j + \text{h.c.} \right) \\ &+ H_1^{i*} m_1^2 H_1^i + H_2^{i*} m_2^2 H_2^i \\ &+ \tilde{\mathbf{q}}_L^{i*} \mathbf{M}_Q^2 \tilde{\mathbf{q}}_L^i + \tilde{\mathbf{l}}_L^{i*} \mathbf{M}_L^2 \tilde{\mathbf{l}}_L^i + \tilde{\mathbf{u}}_R^* \mathbf{M}_U^2 \tilde{\mathbf{u}}_R + \tilde{\mathbf{d}}_R^* \mathbf{M}_D^2 \tilde{\mathbf{d}}_R + \tilde{\mathbf{e}}_R^* \mathbf{M}_E^2 \tilde{\mathbf{e}}_R \\ &+ \frac{1}{2} M_1 \tilde{B} \tilde{B} + \frac{1}{2} M_2 \left( \tilde{W}^3 \tilde{W}^3 + 2\tilde{W}^+ \tilde{W}^- \right) + \frac{1}{2} M_3 \tilde{g} \tilde{g}. \end{aligned} \tag{3}$$

Here $i$ and $j$ are SU(2) indices ($\epsilon_{12} = +1$), $\mathbf{Y}$'s, $\mathbf{A}$'s and $\mathbf{M}$'s are $3 \times 3$ matrices in generation space, and the other boldface letter are vectors in generation space.

The one-loop effective potential for the Higgs fields in the dimensional reduction renormalization scheme then follows (see [10]):

$$\begin{aligned} V &= \mu_1^2 H_1^{i*} H_1^i + \mu_2^2 H_2^{i*} H_2^i - B\mu \left( \epsilon_{ij} H_1^i H_2^j + \text{h.c.} \right) \\ &+ \frac{1}{8}(g^2 + g'^2) \left( H_1^{i*} H_1^i - H_2^{i*} H_2^i \right)^2 + \frac{1}{2} g^2 \left| H_1^{i*} H_2^i \right|^2 + \Delta V_1, \end{aligned} \tag{4}$$

with $\mu_1^2 = \mu^2 + m_1^2$, $\mu_2^2 = \mu^2 + m_2^2$ and the one-loop corrections

$$\Delta V_1 = \frac{1}{64\pi^2} \text{Str} \left[ \mathcal{M}^4 \left( \log \frac{\mathcal{M}^2}{Q^2} - \frac{3}{2} \right) \right]. \tag{5}$$

The supertrace is defined as $\text{Str} f(\mathcal{M}^2) = \sum_i C_i (-1)^{2s_i} (2s_i + 1) f(m_i^2)$ where $C_i$ is the color degrees of freedom and $s_i$ is the spin of the $i^{th}$ particle.

Electroweak symmetry breaking is caused by both $H_1^1$ and $H_2^2$ acquiring vacuum expectation values,

$$\langle H_1^1 \rangle = v_1, \qquad \langle H_2^2 \rangle = v_2, \tag{6}$$

with $g^2(v_1^2 + v_2^2) = 2m_W^2$, with the further assumption that vacuum expectation values of all other scalar fields (in particular, squark and sleptons) vanish. This avoids color and/or charge breaking vacua.

At the tree level, the correct symmetry breaking pattern results if

$$\mu_1^2 + \mu_2^2 > 2|B\mu|, \tag{7}$$

to guarantee a potential bounded from below, and

$$\mu_1^2 \mu_2^2 < B^2 \mu^2, \tag{8}$$

to have a saddle point and not a minimum at $v_1 = v_2 = 0$. At the one-loop level, the potential is automatically bounded from below thanks to the logarithms in the one-loop contributions, and the origin is not an extremum.

The minimization conditions of the potential (4) can be written as

$$\begin{aligned} 0 &= \sqrt{2}v \left[ \mu_1^2 \cos^2\beta - \mu_2^2 \sin^2\beta + \tfrac{1}{8}(g^2 + g'^2)v^2 \cos 2\beta \right] + \Delta T_1 \ , \\ 0 &= \sqrt{2}v \left[ \tfrac{1}{2}(\mu_1^2 + \mu_2^2)\sin 2\beta - B\mu \right] + \Delta T_2 \ , \end{aligned} \tag{9}$$

where $v^2 = v_1^2 + v_2^2$, $\tan\beta = v_2/v_1$ and $\Delta T_1$, $\Delta T_2$ are one-loop tadpole contributions [10]. These minimization conditions allow one to trade two of the Higgs potential parameters $\mu_1^2$, $\mu_2^2$ and $B\mu$ for the $Z$ boson mass $m_Z^2 = \tfrac{1}{2}(g^2 + g'^2)(v_1^2 + v_2^2)$ and the ratio of vevs $\tan\beta$. The third parameter can further be reexpressed in terms of the mass of one of the physical Higgs bosons.

When diagonalizing the mass matrix for the scalar Higgs fields, besides a charged and a neutral would-be Goldstone bosons which become the longitudinal polarizations of the $W^\pm$ and $Z$ gauge bosons, one finds a neutral CP-odd Higgs boson $A$, two neutral CP-even Higgs bosons $H_{1,2}$ and a charged Higgs boson $H^\pm$, which will play an important role in the calculation. Choosing as independent parameter the mass $m_A$ of the CP-odd Higgs boson, the masses of the other Higgs bosons are given by

$$\mathcal{M}_H^2 = \begin{pmatrix} m_A^2 \cos^2\beta + m_Z^2 \sin^2\beta + \Delta\mathcal{M}_{11}^2 & -\sin\beta\cos\beta(m_A^2 + m_Z^2) + \Delta\mathcal{M}_{12}^2 \\ -\sin\beta\cos\beta(m_A^2 + m_Z^2) + \Delta\mathcal{M}_{21}^2 & m_A^2 \sin^2\beta + m_Z^2 \cos^2\beta + \Delta\mathcal{M}_{22}^2 \end{pmatrix} \tag{10}$$

$$m_{H^\pm}^2 = m_A^2 + m_W^2 + \Delta_\pm. \tag{11}$$

The quantities $\Delta\mathcal{M}_{ij}^2$ and $\Delta_\pm$ are the one-loop radiative corrections coming from virtual (s)top and (s)bottom loops, calculated within the effective potential approach as in [11]. Diagonalization of $\mathcal{M}_H^2$ gives the two CP-even Higgs boson masses, $m_{H_{1,2}}$, and their mixing angle $\alpha$ $(-\pi/2 < \alpha < 0)$.

The neutralinos $\tilde{\chi}_i^0$ are linear combination of the neutral gauge bosons $\tilde{B}$, $\tilde{W}_3$ and of the neutral higgsinos $\tilde{H}_1^0$, $\tilde{H}_2^0$. In this basis, their mass matrix

$$\mathcal{M}_{\tilde{\chi}_{1,2,3,4}^0} = \begin{pmatrix} M_1 & 0 & -\frac{g'v_1}{\sqrt{2}} & +\frac{g'v_2}{\sqrt{2}} \\ 0 & M_2 & +\frac{gv_1}{\sqrt{2}} & -\frac{gv_2}{\sqrt{2}} \\ -\frac{g'v_1}{\sqrt{2}} & +\frac{gv_1}{\sqrt{2}} & 0 & -\mu \\ +\frac{g'v_2}{\sqrt{2}} & -\frac{gv_2}{\sqrt{2}} & -\mu & 0 \end{pmatrix} \tag{12}$$

is diagonalized analytically to give four neutral Majorana states,

$$\tilde{\chi}_i^0 = Z_{i1}\tilde{B} + Z_{i2}\tilde{W}^3 + Z_{i3}\tilde{H}_1^0 + Z_{i4}\tilde{H}_2^0, \tag{13}$$

the lightest of which, to be called $\chi$, is then the candidate for the particle making up the dark matter in the universe.

The charginos are linear combinations of the charged gauge bosons $\tilde{W}^\pm$ and of the charged higgsinos $\tilde{H}_1^-$, $\tilde{H}_2^+$. Their mass mass terms are given by

$$\begin{pmatrix} \tilde{W}^- & \tilde{H}_1^- \end{pmatrix} \mathcal{M}_{\tilde{\chi}^\pm} \begin{pmatrix} \tilde{W}^+ \\ \tilde{H}_2^+ \end{pmatrix} + \text{h.c.} \tag{14}$$

Their mass matrix,

$$\mathcal{M}_{\tilde{\chi}^\pm} = \begin{pmatrix} M_2 & gv_2 \\ gv_1 & \mu \end{pmatrix}, \tag{15}$$

is diagonalized by the following linear combinations

$$\tilde{\chi}_i^- = U_{i1}\tilde{W}^- + U_{i2}\tilde{H}_1^-, \tag{16}$$
$$\tilde{\chi}_i^+ = V_{i1}\tilde{W}^+ + V_{i2}\tilde{H}_1^+. \tag{17}$$

We choose $\det(U) = 1$ and $U^*\mathcal{M}_{\tilde{\chi}^\pm}V^\dagger = \text{diag}(m_{\tilde{\chi}_1^\pm}, m_{\tilde{\chi}_2^\pm})$ with non-negative chargino masses $m_{\tilde{\chi}_i^\pm} \geq 0$.

When discussing the squark mass matrix including mixing, it is convenient to choose a basis where the squarks are rotated in the same way as the corresponding quarks in the standard model. We follow the conventions of the particle data group [12] and put the mixing in the left-handed $d$-quark fields, so that the definition of the Cabibbo-Kobayashi-Maskawa matrix is $\mathbf{K} = \mathbf{V}_1\mathbf{V}_2^\dagger$, where $\mathbf{V}_1$ ($\mathbf{V}_2$) rotates the interaction left-handed $u$-quark ($d$-quark) fields to mass eigenstates. For sleptons we choose an analogous basis, but due to the masslessness of neutrinos no analog of the CKM matrix appears.

We then obtain the general $6 \times 6$ $\tilde{u}$- and $\tilde{d}$-squark mass matrices:

$$\mathcal{M}_{\tilde{u}}^2 = \begin{pmatrix} \mathbf{M}_Q^2 + \mathbf{m}_u^\dagger\mathbf{m}_u + D_{LL}^u\mathbf{1} & -\mathbf{m}_u^\dagger(\mathbf{A}_U^\dagger + \mu^*\cot\beta) \\ -(\mathbf{A}_U + \mu\cot\beta)\mathbf{m}_u & \mathbf{M}_U^2 + \mathbf{m}_u\mathbf{m}_u^\dagger + D_{RR}^u\mathbf{1} \end{pmatrix}, \tag{18}$$

$$\mathcal{M}_{\tilde{d}}^2 = \begin{pmatrix} \mathbf{K}^\dagger\mathbf{M}_Q^2\mathbf{K} + \mathbf{m}_d\mathbf{m}_d^\dagger + D_{LL}^d\mathbf{1} & -\mathbf{m}_d^\dagger(\mathbf{A}_D^\dagger + \mu^*\tan\beta) \\ -(\mathbf{A}_D + \mu\tan\beta)\mathbf{m}_d & \mathbf{M}_D^2 + \mathbf{m}_d^\dagger\mathbf{m}_d + D_{RR}^d\mathbf{1} \end{pmatrix}, \tag{19}$$

and the general sneutrino and charged slepton masses

$$\mathcal{M}_{\tilde{\nu}}^2 = \mathbf{M}_L^2 + D_{LL}^\nu\mathbf{1} \tag{20}$$

$$\mathcal{M}_{\tilde{e}}^2 = \begin{pmatrix} \mathbf{M}_L^2 + \mathbf{m}_e\mathbf{m}_e^\dagger + D_{LL}^e\mathbf{1} & -\mathbf{m}_e^\dagger(\mathbf{A}_E^\dagger + \mu^*\tan\beta) \\ -(\mathbf{A}_E + \mu\tan\beta)\mathbf{m}_e & \mathbf{M}_E^2 + \mathbf{m}_e^\dagger\mathbf{m}_e + D_{RR}^e\mathbf{1} \end{pmatrix}. \tag{21}$$

Here

$$D_{LL}^f = m_Z^2\cos 2\beta(T_{3f} - e_f\sin^2\theta_w), \tag{22}$$

$$D^f_{RR} = m_Z^2 \cos 2\beta e_f \sin^2 \theta_w. \tag{23}$$

In the chosen basis, $\mathbf{m}_u = \mathrm{diag}\,(m_\mathrm{u}, m_\mathrm{c}, m_\mathrm{t})$, $\mathbf{m}_d = \mathrm{diag}\,(m_\mathrm{d}, m_\mathrm{s}, m_\mathrm{b})$ and $\mathbf{m}_e = \mathrm{diag}(m_e, m_\mu, m_\tau)$.

The slepton and squark mass eigenstates $\tilde{f}_k$ ($\tilde{\nu}_k$ with $k = 1, 2, 3$ and $\tilde{e}_k$, $\tilde{u}_k$ and $\tilde{d}_k$ with $k = 1, \ldots, 6$) diagonalize the previous mass matrices and are related to the current sfermion eigenstates $\tilde{\mathbf{f}}_L$ and $\tilde{\mathbf{f}}_R$ via ($a = 1, 2, 3$)

$$\tilde{f}_{La} = \sum_{k=1}^{6} \tilde{f}_k \Gamma^{*ka}_{FL}, \tag{24}$$

$$\tilde{f}_{Ra} = \sum_{k=1}^{6} \tilde{f}_k \Gamma^{*ka}_{FR}. \tag{25}$$

The squark and charged slepton mixing matrices $\boldsymbol{\Gamma}_{UL,R}$, $\boldsymbol{\Gamma}_{DL,R}$ and $\boldsymbol{\Gamma}_{EL,R}$ have dimension $6 \times 3$, while the sneutrino mixing matrix $\boldsymbol{\Gamma}_{\nu L}$ has dimension $3 \times 3$.

For simplicity, and to get in touch with published papers [3], we then make a simple ansatz for the up-to-now arbitrary soft supersymmetry-breaking parameters:

$$\begin{aligned} \mathbf{A}_U &= \mathrm{diag}(0, 0, A_t) \\ \mathbf{A}_D &= \mathrm{diag}(0, 0, A_b) \\ \mathbf{A}_E &= 0 \\ \mathbf{M}_Q &= \mathbf{M}_U = \mathbf{M}_D = \mathbf{M}_E = \mathbf{M}_L = m_0 \mathbf{1}. \end{aligned} \tag{26}$$

This allows the squark mass matrices to be diagonalized analytically. For example, for the top squark one has, in terms of top squark mixing angle $\theta_{\tilde{t}}$,

$$\Gamma^{\tilde{t}_1 \tilde{t}}_{UL} = \Gamma^{\tilde{t}_2 \tilde{t}}_{UR} = \cos\theta_{\tilde{t}}, \qquad \Gamma^{\tilde{t}_2 \tilde{t}}_{UL} = -\Gamma^{\tilde{t}_1 \tilde{t}}_{UR} = \sin\theta_{\tilde{t}}. \tag{27}$$

Notice that the ansatz (26) implies the absence of tree-level flavor changing neutral currents in all sectors of the model. It is not however the more general ansatz for the absence of tree-level FCNC's, which would only demand that the trilinear couplings $\mathbf{A}_i$, the soft mass matrices $\mathbf{M}_i^2$ and also $\mathbf{K}^\dagger \mathbf{M}_Q^2 \mathbf{K}$ be diagonal, and not necessarily equal to each other. Notice also that this is typically not what is obtained in low-energy supergravity models with a universal scalar mass at the grand-unification (or Planck) scale, in which the running of the scalar masses down to the electroweak scale generates off-diagonal terms and tree-level FCNC's in the squark sector.

## The $\mathrm{b} \to \mathrm{s}\gamma$ branching ratio

In the calculation of the branching ratio for $\mathrm{b} \to \mathrm{s}\gamma$ we follow [13]. In the standard model, the rate is given by

$$\Gamma(\mathrm{b} \to \mathrm{s}\gamma) = \frac{m_b^5 |A_W|^2}{16\pi}, \tag{28}$$

with

$$A_W = \frac{3eg^2}{32\pi^2 m_W^2} K^*_{\mathrm{ts}} K_{\mathrm{tb}}\, x_{tW} f_1(x_{tW}), \tag{29}$$

where $e$ and $g$ are evaluated at the b mass scale, and the function $f_1$ arises from the loop integration:
$$f_1(x) = \frac{8x^3 - 3x^2 - 12x + 7 + 6x(2 - 3x)\log(x)}{36(x-1)^4}. \tag{30}$$

Here and in the following $x_{ij} = m_i^2/m_j^2$.

For a general supersymmetric model, contributions from charged Higgs bosons $H^-$, charginos $\tilde{\chi}^-$, gluinos $\tilde{g}$ and neutralinos $\tilde{\chi}^0$ have to be added to $A_W$. In our case, only the standard model contribution $A_W$ and those for $H^-$ and $\tilde{\chi}^-$ enter, since the gluino and neutralino contributions vanish because of our ansatz for the squark mass matrices (absence of tree-level FCNC's).

$A_{H^-}$ and $A_{\tilde{\chi}^-}$ can be written as (see, e.g. [13])

$$A_{H^-} = \frac{eg^2}{32\pi^2 m_W^2} K_{\text{ts}}^* K_{\text{tb}} \, x_{tH} \left[\cot^2\beta \, f_1(x_{tH}) + f_2(x_{tH})\right], \tag{31}$$

$$A_{\tilde{\chi}^-} = -\sum_{c=1}^{2}\sum_{k=1}^{6} \frac{e}{16\pi^2 m_{\tilde{\chi}_c^\pm}^2} \bigg[g_{L\tilde{u}_k\tilde{\chi}_c^\pm \text{b}} g^*_{L\tilde{u}_k\tilde{\chi}_c^\pm \text{s}} \, f_1(x_{\tilde{\chi}_c^-\tilde{u}_k}) + \\ + g_{R\tilde{u}_k\tilde{\chi}_c^\pm \text{b}} g^*_{L\tilde{u}_k\tilde{\chi}_c^\pm \text{s}} \frac{m_{\tilde{\chi}_c^-}}{m_b} f_3(x_{\tilde{\chi}_c^-\tilde{u}_k})\bigg], \tag{32}$$

where $g_{L\tilde{u}_k\tilde{\chi}_c^\pm d}$ and $g_{R\tilde{u}_k\tilde{\chi}_c^\pm d}$ are the left- and right-handed down-quark–chargino–up-squark vertex constants given by

$$g_{L\tilde{u}_k\tilde{\chi}_c^\pm d} = \sum_{i=1}^{3} \left(V_{c2}^* \Gamma_{UR}^{ki} Y_U^i - gV_{c1}^* \Gamma_{UL}^{ki}\right) K_{id} \tag{33}$$

$$g_{R\tilde{u}_k\tilde{\chi}_c^\pm d} = \sum_{i=1}^{3} U_{c2} \Gamma_{UL}^{ki} K_{id} Y_D^d. \tag{34}$$

The additional loop integration functions entering are

$$f_2(x) = \frac{5x^2 - 8x + 3 + 2(2 - 3x)\log(x)}{6(x-1)^3} \tag{35}$$

and

$$f_3(x) = \frac{-7x^2 + 12x - 5 - 2x(2 - 3x)\log(x)}{6(x-1)^3}. \tag{36}$$

It is important for our result that the loop integration functions $f_i(x)$ ($i = 1, 2, 3$) are decreasing positive functions of $x$ [14].

The b $\to$ s$\gamma$ branching ratio is finally evaluated as

$$\text{BR}(\text{b} \to \text{s}\gamma) = \text{BR}(\text{b} \to \text{c}e\bar{\nu}) \frac{\Gamma(\text{b} \to \text{s}\gamma)}{\Gamma(\text{b} \to \text{c}e\bar{\nu})} \tag{37}$$

with

$$\text{BR}(\text{b} \to \text{c}e\bar{\nu}) = 0.107 \tag{38}$$

and
$$\Gamma(\text{b} \to \text{c}e\bar{\nu}) = \frac{m_\text{b}^5 G_F^2}{192\pi^3} |K_{\text{cb}}|^2 \rho(x_{cb}) \lambda(m_b, x_{cb}), \tag{39}$$

where
$$\rho(x) = 1 - 8x^2 + 8x^6 - x^8 - 24x^4 \log(x) \tag{40}$$

is a phase space factor, and
$$\lambda(m, x) = 1 - \frac{2}{3\pi} \alpha_s(m) f_S(x) \tag{41}$$

is a QCD correction factor for the semileptonic decay ($f_S(x_{cb}) \simeq 2.4$ [15]).

A complication in the analysis is given by possibly large QCD corrections to the b $\to$ s$\gamma$ decay, which at present are plagued by theoretical uncertainties (see e.g. ref. [16]). Therefore we preferentially opted for the tree-level value described in this section. In the discussion we will comment on the effect of including QCD corrections according to the procedure outlined in ref. [13].

## The direct detection rate

The rate for direct detection of galactic neutralinos, integrated over deposited energy assuming no energy threshold, is
$$R = \sum_i N_i n_\chi \langle \sigma_{i\chi} v \rangle, \tag{42}$$

where $N_i$ is the number of nuclei of species $i$ in the detector, $n_\chi$ is the local galactic neutralino number density, $\sigma_{i\chi}$ is the neutralino-nucleus elastic cross section, and the angular brackets denote an average over $v$, the neutralino speed relative to the detector.

We take the local galactic neutralino velocity distribution as a truncated gaussian, which in the detector frame moving at speed $v_O$ relative to the galactic halo reads
$$f(v) = \frac{1}{\mathcal{N}_\text{cut}} \frac{v^2}{uv_O\sigma} \left\{ \exp\left[-\frac{(u-v_O)^2}{2\sigma^2}\right] - \exp\left[-\frac{\min(u+v_O, v_\text{cut})^2}{2\sigma^2}\right] \right\} \tag{43}$$

for $v_\text{esc} < v < \sqrt{v_\text{esc}^2 + (v_O + v_\text{cut})^2}$ and zero otherwise, with $u = \sqrt{v^2 + v_\text{esc}^2}$ and
$$\mathcal{N}_\text{cut} = \frac{v_\text{cut}}{\sigma} \exp\left(-\frac{v_\text{cut}^2}{2\sigma^2}\right) - \sqrt{\frac{\pi}{2}} \text{erf}\left(\frac{v_\text{cut}}{\sqrt{2}\sigma}\right). \tag{44}$$

Numerically, we have taken the halo line-of-sight velocity dispersion $\sigma =$120 km/s, the galactic escape speed $v_\text{cut} = 600$ km/s, the relative Earth-halo speed $v_O = 264$ km/s (a yearly average) and the Earth escape speed $v_\text{esc} = 11.9$ km/s. We have adopted a local dark matter density $m_\chi n_\chi = 0.3$ GeV/cm$^3$ whenever the calculated neutralino relic density $\Omega_\chi h^2 > \Omega_\text{gal.DM} h^2$, the density of dark matter in galactic halos averaged over the entire universe. This is meant to represent the minimum value for which neutralinos could make up the totality of the galactic dark matter. When $\Omega_\chi h^2 < \Omega_\text{gal.DM} h^2$, we have scaled $n_\chi$ proportionally to $\Omega_\chi/\Omega_{\rm gal.DM}$. The value to choose for $\Omega_{\rm gal.DM}h^2$ is quite uncertain, both because of uncertainties in the density and extension of galactic halos and because of the poorly known relation between the universally-averaged and the local dark matter densities. Just for comparison, we have chosen the same value $\Omega_{\rm gal.DM}h^2 = 0.025$ as in ref. [3], but we call to the attention of the reader that values smaller by one order of magnitude would still be acceptable.

The neutralino-nucleus elastic cross section can be written as

$$\sigma_{i\chi} = \frac{1}{4\pi v^2} \int_0^{4m_{i\chi}^2 v^2} {\rm d}q^2 G_{i\chi}^2(q^2), \qquad (45)$$

where $m_{i\chi}$ is the neutralino-nucleus reduced mass, $q$ is the momentum transfer and $G_{i\chi}(q^2)$ is the effective neutralino-nucleus vertex. In this calculation we write

$$G_{i\chi}^2(q^2) = A_i^2 F_S^2(q^2) G_S^2 + 4\Lambda_i^2 F_A^2(q^2) G_A^2, \qquad (46)$$

(often $\Lambda_i^2$ appears as $\lambda^2 J(J+1)$ ) and assume gaussian nuclear form factors [17]

$$F_S(q^2) = F_A(q^2) = \exp(-q^2 R_i^2/6\hbar^2), \qquad (47)$$

$$R_i = (0.3 + 0.89 A_i^{1/3}){\rm fm}, \qquad (48)$$

which should provide us with a good approximation of the integrated detection rate [18], in which we are only interested. Since the non-zero-spin nuclei we consider have an unpaired proton, we consider the scalar and axial neutralino-proton vertices $G_S$ and $G_A$, a neutron admixture giving a contribution that can be neglected. Using heavy-squark effective lagrangians [19], we get

$$G_S = \sum_{q=u,d,s,c,b,t} \langle \bar{q}q \rangle \left( \sum_{h=H_1,H_2} \frac{g_{h\chi\chi} g_{hqq}}{m_h^2} - \frac{1}{2} \sum_{k=1}^{6} \frac{g_{L\tilde{q}_k \chi q} g_{R\tilde{q}_k \chi q}}{m_{\tilde{q}_k}^2} \right) \qquad (49)$$

and

$$G_A = \sum_{q=u,d,s} \Delta q \left( \frac{g_{Z\chi\chi} g_{Zqq}}{m_Z^2} + \frac{1}{8} \sum_{k=1}^{6} \frac{g_{L\tilde{q}_k \chi q}^2 + g_{R\tilde{q}_k \chi q}^2}{m_{\tilde{q}_k}^2} \right). \qquad (50)$$

The $g$'s are elementary vertices involving the particles indicated by the indices, and they read

$$g_{h\chi\chi} = \begin{cases} (gZ_{\chi 2} - g_y Z_{\chi 1})(-Z_{\chi 3}\cos\alpha + Z_{\chi 4}\sin\alpha), & \text{for } H_1, \\ (gZ_{\chi 2} - g_y Z_{\chi 1})(Z_{\chi 3}\sin\alpha + Z_{\chi 4}\cos\alpha), & \text{for } H_2, \end{cases} \qquad (51)$$

$$g_{hqq} = \begin{cases} -Y_q \cos\alpha/\sqrt{2}, & \text{for } H_1, \\ +Y_q \sin\alpha/\sqrt{2}, & \text{for } H_2, \end{cases} \qquad (52)$$

$$g_{Z\chi\chi} = \frac{g}{2\cos\theta_W}\left(Z_{\chi 3}^2 - Z_{\chi 4}^2\right) \qquad (53)$$

$$g_{Zqq} = -\frac{g}{2\cos\theta_W} T_{3q}, \qquad (54)$$

$$g_{L\tilde{q}_k \chi q} = g_{LL}\Gamma_{QL}^{kq} + g_{RL}\Gamma_{QR}^{kq}, \qquad (55)$$

$$g_{R\tilde{q}_k \chi q} = g_{LR}\Gamma_{QL}^{kq} + g_{RR}\Gamma_{QR}^{kq}, \qquad (56)$$

with

$$g_{LL} = -\frac{1}{\sqrt{2}}\left(T_{3q}gZ_{\chi 2} + \frac{1}{3}g_y Z_{\chi 1}\right), \quad (57)$$

$$g_{RR} = \sqrt{2}e_q g_y Z_{\chi 1}, \quad (58)$$

$$g_{LR} = g_{RL} = \begin{cases} -Y_q Z_{\chi 3}, & \text{for } q = \text{u, c, t,} \\ -Y_q Z_{\chi 4}, & \text{for } q = \text{d, s, b,} \end{cases} \quad (59)$$

and

$$Y_q = \begin{cases} m_q/v_2, & \text{for } q = \text{u, c, t,} \\ m_q/v_1, & \text{for } q = \text{d, s, b.} \end{cases} \quad (60)$$

Numerically, we have taken [20]

$$m_\text{u}\langle \bar{\text{u}}\text{u}\rangle = 0.023 m_\text{p}, \qquad m_\text{d}\langle \bar{\text{d}}\text{d}\rangle = 0.034 m_\text{p}, \quad (61)$$

$$m_\text{s}\langle \bar{\text{s}}\text{s}\rangle = 0.14 m_\text{p}, \qquad m_\text{c}\langle \bar{\text{c}}\text{c}\rangle = m_\text{b}\langle \bar{\text{b}}\text{b}\rangle = m_\text{t}\langle \bar{\text{t}}\text{t}\rangle = 0.595 m_\text{p}, \quad (62)$$

and [21]

$$\Delta\text{u} = 0.77, \qquad \Delta\text{d} = -0.49, \qquad \Delta\text{s} = -0.15. \quad (63)$$

Moreover, we have used

$$\Lambda_\text{Al}^2 = 0.35, \qquad \Lambda_\text{Na}^2 = 0.041 \quad \text{and} \quad \Lambda_\text{I}^2 = 0.007, \quad (64)$$

according to the odd-group model [22].

One should be aware that both our choice of nuclear form factors and of neutralino-nucleon vertices and the numerical values adopted for the nucleon matrix elements are at best approximate. A more sophisticated treatment (see discussion and references in [23]), would however, beyond still presenting theoretical and calculational weaknesses, change the rate values by much less than the spread due to the unknown supersymmetric parameters.[1]

## Sampling of supersymmetric parameter space

When solving the minimal supersymmetric standard models defined above we let the universal (weak-scale) sfermion mass parameter $m_0$ vary between 100 and 3000 GeV, the soft supersymmetry breaking parameters $A_b$ and $A_t$ between $-3m_0$ and $3m_0$, $\tan\beta$ between 1.2 and 50, the pseudoscalar mass $m_A$ between its experimentally allowed lower bound and 1000 GeV, and we fix the top quark mass at $m_t = 175$ GeV.

For the model scan, two alternative sets of parameters are used: (1) the usual $\mu$ and $M_2$, both varied logarithmically in the interval $[-5000, 5000]$ GeV, and (2) the mass of the lightest neutralino $m_\chi$ and its gaugino fraction $Z_g = Z_{\chi 1}^2 + Z_{\chi 2}^2$, varied logarithmically in $[-5000, 5000]$ GeV and linearly in $[0.00001, 0.99999]$. To start with, we enforce the gaugino mass unification conditions (1).

---

[1] The b → sγ formulas in ref. [23] and in the accompanying computer code neutdriver seem to contain some errors. After correction and fixing of some bugs, neutdriver gave results in qualitative agreement with ours. We thank G. Jungman for providing us with the source code.

With this definition of the MSSM models (which we note is the same as in [3]), we generate model parameters randomly within the bounds mentioned above and proceed to calculate the mass spectrum, the couplings, the Z-boson invisible width, etc. We keep only models that satisfy the accelerator constraints given in the 1995 Review of Particle Properties [12]. In addition, we drop models that violate the 95% C.L. limits from the CLEO experiment [1] $1.0 \cdot 10^{-4} < \text{BR}(\text{b} \to \text{s}\gamma) < 3.4 \cdot 10^{-4}$. We ask for 4,000 models satisfying the accelerator constraints including $\text{b} \to \text{s}\gamma$.

Then for each model allowed by the accelerator constraints we calculate the relic density of neutralinos $\Omega_\chi h^2$. We use the formalism in ref. [24] to carefully treat resonant annihilations and threshold effects, keeping finite widths of unstable particles, including all two-body annihilation channels of neutralinos. The annihilation cross sections used were derived using a novel helicity projection technique [25], and were checked against published results for several of the subprocesses. Only models that would not overclose the universe, i.e. in which $\Omega_\chi h^2 < 1$, are considered cosmologically viable.

In fig. 1 we show the $\mu$–$M_2$ location of the viable models that survive all experimental bounds, including the $\Omega_\chi h^2$ and $\text{b} \to \text{s}\gamma$ constraints. Fig. 1(a) refers to the $\mu$–$M_2$ sampling, and fig. 1(b) to the $m_\chi$–$Z_g$ sampling.

At this point we compute integrated direct detection rates for pure germanium ($^{76}$Ge), sapphire ($\text{Al}_2\text{O}_3$) and sodium iodide (NaI) detectors, which are representative of devices presently under research or development. The rates resulting in experimentally and cosmologically viable models are presented in fig. 2 for a Ge detector. Again, fig. 2(a) refers to the $\mu$–$M_2$ sampling, and fig. 2(b) to the $m_\chi$–$Z_g$ sampling.

The huge spread of possible rates, even at fixed neutralino mass, is evident. This hinders the predictability of the class of MSSM models we consider, and is one of the driving reasons for introducing more restrictive, and so more predictive, relations among the model parameters.

A fact should however be noticed: in the two samplings, *no* points have interesting detection rates, say above 1 event/kg/day in Ge. Can we conclude that in the class of models we consider the $\text{b} \to \text{s}\gamma$ constraint is so strong to exclude interesting detection rates?

Compare figs. 2(a) and 2(b): the aesthetic appearance of the clouds of points is quite different. This is only due to the different *a priori* probabilities used in the two samplings. It is obvious that Nature, if she has chosen supersymmetry, has realized just one of the models appearing as dots in our figures. By the same token, we have no reason to prefer an *a priori* probability measure over another. And so we must not be lead astray by attaching a probability to the points in the figures. We must not turn sentences like 'most of the models give low detection rates' into statements on the relative likelihood of high and low detection rates. The same applies of course to histograms derived from such samplings. And if comparison of figs. 2(a) and 2(b) is not convincing enough, it should become inescapable in the following.

We perform two special scans, of 500 models each, demanding to consider only models in which the scalar piece of the neutralino-proton cross section is larger than a tenth of the corresponding Dirac neutrino-proton cross section. This in order to pick out points with the highest detection rates. In the first scan we sample in $\mu$–$M_2$ as before but restrict the range of the Higgs pseudoscalar mass to $m_A \in [0, 60]\,\mathrm{GeV}$. In the second scan, we sample in $m_\chi$–$Z_g$ with the restricted $m_A$ range and further demand $m_\chi \in [800, 1200]\,\mathrm{GeV}$ and $Z_g \in [0.01, 0.99]$. The results of these special scans for Ge, $\mathrm{Al}_2\mathrm{O}_3$ and NaI are shown in fig. 3, combined with (the top parts of) those of the previous samplings. Remarkably, the high-rate zones, empty before, are now filled with points. Particularly striking is the concentration of points around $m_\chi \simeq 1000\,\mathrm{GeV}$, which obviously comes from the second special sampling. In fig. 4 we also show the dependence of the counting rates on the gaugino fraction $Z_g$ for the combined sample. The upper band, corresponding to the two special scans, is almost flat. The slightly higher rates at $Z_g > 0.5$ are essentially due to a weaker or even absent $\Omega$-suppression of the galactic neutralino density.

We remind that all points shown in figs. 3 and 4 are compatible with the experimental and cosmological constraints mentioned above, which include the b $\to$ s$\gamma$ bounds. And so we conclude that, as far as only these constraints are considered, there are viable models with an integrated counting rate as large as 10 events/kg/day in Ge, and even 100 events/kg/day in NaI. These models might well be already probed (and excluded) by current dark matter searches.

## Discussion

As we have shown, contrary to the results of [3], we find models with as high Ge counting rate as 10 events/kg/day which still do not violate the b $\to$ s$\gamma$ bound.

In fig. 5 we plot the b $\to$ s$\gamma$ rate as a function of the $H^\pm$ mass. The Ge detection rate decreases from the upper left to the lower right panel. The vertical band of points at relatively low $m_{H^\pm}$ belongs to the two special samples, and shows best at high Ge rates. The curve at relatively large $m_{H^\pm}$, most pronounced at very low Ge rates, comes from the naive $\mu$–$M_2$ and $m_\chi$–$Z_g$ samplings combined, and describes the naively-expected relationship between detection rate and charged Higgs boson mass. The two groups of points are detached, but this is only an artifact of the sampling procedure.

The reason for having an acceptable b $\to$ s$\gamma$ from a relatively light (100 - 200 GeV) $H^\pm$ happens to be a cancellation between the charged Higgs contribution and the contribution from the lighter of the two charginos [14]. The latter may be sizable at large $\tan\beta$ when the top-squark mixing is substantial and the lightest chargino is dominantly a charged higgsino $\tilde\chi_2^\pm \approx \tilde H^\pm$ and is relatively light ($|M_2| \gg |\mu| \gtrsim m_W$). Under these conditions, the dominant chargino contribution is well approximated by

$$A_{\tilde\chi^-} \approx -\mathrm{sign}(\mu\theta_{\tilde t}) \frac{eg^2}{16\pi^2 m_W^2} K_{ts}^* K_{tb} \frac{m_W}{\sqrt{2}|m_{\tilde\chi_2^-}|\cos\beta} \cdot$$
$$\cdot \left|U_{\tilde\chi_2^- \tilde H^-} V_{\tilde\chi_2^+ \tilde H^+} \sin\theta_{\tilde t} \cos\theta_{\tilde t}\right| \left|f_3(x_{\tilde t_1 \tilde\chi_2^-}) - f_3(x_{\tilde t_2 \tilde\chi_2^-})\right|. \qquad (65)$$

This charged higgsino contribution can be negative and effectively cancel the W boson and charged Higgs boson contributions, which are always positive, when the top squark mixing angle $\theta_{\tilde t}$ and the parameter $\mu$ have the same sign. In terms of the soft-supersymmetry breaking parameters this amounts to

$$\mu(A_t + \mu \cot \beta) < 0, \qquad (66)$$

or in common instances in which $A_t = \mathcal{O}(m_0) \gg \mu/\tan\beta$ to roughly $\mu A_t < 0$. In our class of models it is possible to satisfy the previous condition (66) for both $\mu$ positive and $\mu$ negative, because we are free to choose the sign and magnitude of $A_t$. This freedom is lost in models imposing additional theoretical constraints, for example in no-scale models or in models with a flat Kähler manifold ($A = 0$ or $A = B - m$ respectively at the unification scale).

We have also performed an analysis including QCD corrections to the b $\rightarrow$ s$\gamma$ amplitudes according to the prescription in ref. [13]. For a given model, the BR(b $\rightarrow$ s$\gamma$) is generally larger than the tree level value, so some of the models that were previous viable have too large a QCD-corrected BR(b $\rightarrow$ s$\gamma$). However, other models that had too low a tree-level b $\rightarrow$ s$\gamma$ decay rate become viable when the QCD corrections increase BR(b $\rightarrow$ s$\gamma$). So our conclusions on the existence of models with high counting rates and acceptable BR(b $\rightarrow$ s$\gamma$) remain valid even after including QCD corrections.

We already mentioned that we have chosen $\Omega_{\mathrm{gal.DM}} h^2 = 0.025$ just for the sake of comparison with ref. [3], and that the actual value to use might even one order of magnitude larger or smaller. Were it smaller, many models in which the detection rate is suppressed just because of a too small relic density would add to the number of models with important detection rates. Fig. 6 shows the predicted rates in Ge versus the calculated neutralino relic density $\Omega_\chi h^2$, for the combined sample of models. The two naive samples fill the triangular shape at the bottom, and the two special samples are the band and cloud in the upper parts. Models to the right of the solid vertical line overclose the universe (and have been plotted to illustrate the trend of $R$ versus $\Omega_\chi h^2$). Rates to the left of the vertical dashed line ($\Omega_\chi = \Omega_{\mathrm{gal.DM}}$) are $\Omega$-suppressed by our simple-minded prescription for the local galactic neutralino density at low $\Omega_\chi$. Lowering $\Omega_{\mathrm{gal.DM}}$ shifts the tip of the 'volcano' upwards and towards the left so that right-hand side continues straight to the left, and raises the cloud of specially-sampled models to still higher rates. This would also lower the typical neutralino mass (here 1000 GeV) for which calculated detection rates are highest. Notice that in some of the interesting models, the neutralino relic density is larger than what is needed for them to fill up galactic halos, i.e. $\Omega_\chi > \Omega_{\mathrm{gal.DM}}$.

Some of the models shown in the figures may indeed be already excluded by current negative searches of halo neutralinos. We have not tagged these models as excluded because a serious analysis would require a more detailed calculation of the predicted rates, involving different differential rates for each experiment, sophisticated nuclear form factors, quenching factors, etc. Such an analysis is out of the scope of the present paper.

We set out to go beyond the restrictive supergravity models and examine neutralino detection rates in a general minimal supersymmetric model. Relaxing our ansatz (26) on the sfermion masses would demand the consideration of several phenomenological bounds from flavor changing neutral currents, as e.g. the $K^0 \bar{K}^0$ and $B^0 \bar{B}^0$ mass differences, the electric dipole moment of the neutron, etc. We have started by relaxing the request of a universal gaugino mass, eqs. (1), inspired by a recent suggestion to reproduce the measured value of $\alpha_s$ at low energies [8]. We have replaced eqs. (1) by $M_1 \simeq 0.3 M_2 \simeq M_3$ and run through our calculation again. The plots we obtain do not differ qualitatively from those we presented, and the quantitative differences are slight. For these reasons, we do not show them here. Our conclusions on the existence of supersymmetric models with acceptable b → sγ branching ratios and important detection rates remain the same.

## Conclusions

In this paper we have shown that even with the same kind of supersymmetric models as in [3], and enforcing the constraints from b → sγ, we obtain a range of predictions for direct dark matter search experiments that are much less restrictive. In particular we have found models in which the integrated counting rates in Ge detectors are calculated to be higher than 10 events/kg/day, and even higher in NaI detectors. The basic reason behind the compatibility of these models with the b → sγ constraints is the non-existence of a lower bound on the charged Higgs boson from b → sγ measurements, since in the presence of a large top squark mixing the $W$ and $H^{\pm}$ contributions to the b → sγ amplitude may be effectively canceled at large $\tan\beta$ by the contribution from a light charged higgsino. This cancellation may occur at both positive and negative values of $\mu$.

We have argued that with the present ignorance of the origin of an effective low-energy supersymmetric theory one should allow the phenomenological parameters to vary over as large a range as possible without imposing unnecessary relations between parameters (but still restricting them according to various experimental bounds coming from particle physics and cosmology). Although we have kept some simplifying assumptions, the large range of the predicted rates is remarkable. As dark matter detectors improve, we expect this range to be successively narrowed.

We have stressed that the density of points in the plots depends on the assumed *a priori* distribution of the parameters. One should not be misled in thinking that rate values in zones where there are more points are more probable than those in which there are few. To illustrate this, we have shown the rates obtained by trading $\mu$ and $M_2$ with the neutralino mass $m_\chi$ and gaugino fraction $Z_g$. The aesthetic appearance of the plots is indeed different. To further stress our point, we have also presented the rates for an *a priori* distribution which privileges light charginos and light charged Higgs bosons. With this choice, high detection rates with acceptable b → sγ branching ratios look "generic." It is therefore apparent that no probability should be attached to the plotted point distributions (or to histograms derived from them), and that the figures can only illustrate possible neutralino detection rates. With this *caveat*, the detection rates are highest for a neutralino mass around 1000 GeV, preferentially more gaugino than higgsino, but these neutralino characteristics are sensitive to the prescription for the Ω-suppression of the galactic neutralino density.

Naturally, one should consider properties that are invariant under an arbitrary change of the *a priori* parameter probabilities, like the maximum (or minimum) values of the quantities of interest (the detection rates in our case). Unfortunately, a thorough and fine scanning of parameter space is computationally very expensive and an alternative analytical extremization seems prohibitively complicated. We therefore have to leave the following question open: are there in fact additional, allowed points in the empty regions of our plots?

## Acknowledgments

This work was supported in part by European Community Twinning and Mobility (Theoretical Astroparticle Network, TAN) grants. L.B. acknowledges the support from the Swedish Natural Science Research Council and wants to thank the Department of Physics at UC Berkeley for hospitality while part of this work was performed. P.G. is supported by the abovementioned TAN and wishes to thank the Department of Physics at Stockholm University and the Laboratoire de Physique Corpusculaire at Collège de France (Paris) for kind hospitality.

# Figure captions

Figure 1: Location of experimentally and cosmologically viable models in the $\mu$–$M_2$ plane for (a) the $\mu$–$M_2$ sample and (b) the $m_\chi$–$Z_g$ sample.

Figure 2: Integrated direct detection rate R off $^{76}$Ge versus neutralino mass $m_\chi$ for (a) the $\mu$–$M_2$ sample and (b) the $m_\chi$–$Z_g$ sample.

Figure 3: Integrated direct detection rate R in $^{76}$Ge, Al$_2$O$_3$ and NaI detectors versus neutralino mass $m_\chi$ for the combined sample.

Figure 4: Integrated direct detection rate R in $^{76}$Ge, Al$_2$O$_3$ and NaI detectors versus neutralino gaugino fraction $Z_g$ for the combined sample.

Figure 5: Branching ratio for b $\to$ s$\gamma$ decays versus charged Higgs boson mass $m_{H^\pm}$ for the combined sample.

Figure 6: Integrated direct detection rate R off $^{76}$Ge versus neutralino relic density $\Omega_\chi h^2$ for the combined sample.